\def\t{e^{{{(r^2 + \frac{\beta^2}{\pi^2}
{\mathrm{sin}}^2(\frac{\pi\tau}{\beta})-R^2)}/\Lambda^2}}+1}
\def\sin{{\mathrm{sin}}}
\def\cos{{\mathrm{cos}}}
\def\s{e^{{{(\rho^2 -R^2)}/\Lambda^2}}+1}
\def\ss{e^{{{(r^2 -R^2)}/\Lambda^2}}+1}
\def\z{{e^{y -y_0} + 1}}
\def\zz{{e^{2(y-y_0)}}}
\def\in{\int_0^\infty}
\def\g{\gamma}
\def\d{\delta}
\def\l{\Lambda}
\def\vp{\varphi}
\def\b{\beta}
\def\ka{\kappa}
\def\arg{\frac{2\pi\tau}{\b}}
\def\be{\begin{equation}}
\def\ee{\end{equation}}
\def\bea{\begin{eqnarray}}
\def\eea{\end{eqnarray}}
\documentclass[12pt]{article}
\title{Analytical approach to the transition to thermal hopping 
in the thin- and thick-wall approximations}
\author{Hatem Widyan\thanks{E--mail : h$\_$widyan@maktoob.com} ,
A. Mukherjee\thanks{E--mail : am@ducos.ernet.in} ,
N. Panchapakesan\thanks{E--mail : panchu@vsnl.com}  
 and R. P. Saxena\thanks{E--mail : rps@ducos.ernet.in} \\ 
	{\em Department of Physics and Astrophysics,} \\
	{\em University of Delhi, Delhi-110 007, India}
	}

\begin {document}
\maketitle
\begin{abstract}
The nature of the transition from the quantum tunneling regime at low
temperatures to the thermal hopping regime at high temperatures is
investigated analytically in scalar field theory. An analytical bounce
solution is presented, which reproduces the action in the thin-wall as
well as thick-wall limits. The transition is first order for the
case of a thin wall while for the thick wall case it is second order.

\end{abstract}
%

%
%
\begin{section}*{I. INTRODUCTION}
\par
 The decay of metastable states is a basic phenomenon of great
generality with numerous applications in a large number of contexts, 
ranging from the decay of the false vacuum \cite{coleman}, e.g., in 
cosmology \cite{linde}, to the creep-type motion of topological
defects in solids \cite{blatter}. At a given temperature $T$ the 
decay rate of a metastable state can be written in the form
$\Gamma= A e^{-S_{E}(T)/\hbar}$, with $S_{E}(T)$ being the 
Euclidean action of the saddle-point configuration (the bounce) and 
$A$ being the prefactor determined by the associated fluctuations.
At zero temperature, the decay is determined by quantum effects.
With increasing temperature, the nature of the decay changes from
quantum to classical. The function $S_{E}(T)$ might either be a smooth
function of temperature or exhibit a kink with a discontinuity
in its derivative at some temperature $T_c$. In the former case, the
transition from the quantum tunneling regime is said to be of second
order while in the latter case it is said to be of first order. The
word transition is appropriate as the crossovers have all the features
of mean-field phase transition upon identifying the Euclidean action
with the free energy \cite{gorokhov}. 

Affleck \cite{affleck} studied the phase transition at nonzero
temperature for a quantum mechanical problem. He argued that a second
order transition from the quantum tunneling regime to the thermal
hopping regime takes place at some critical temperature. Chudnovsky
\cite{chudnovsky} and Garriga \cite{garriga} have given criteria for
determining when the transition is first-order and when it is
second-order. There have been many studies on the order of the phase
transition in the context of condensed matter; for a recent work see
\cite{gorokhov}. On the other hand, there have been very few studies
in quantum field theoretic situations. 

  A field theoretic system, consisting of a single scalar field with a
$\vp$ symmetry breaking term, has been studied recently by
Ferrera \cite{ferrera} and also by us. We have also investigated in
detail the nature of the phase transition for single scalar field
theory with $\vp^3$ symmetry breaking \cite{hatem}. Our numerical
results show that, for small values of the symmetry-breaking coupling
$f$, the transition from the quantum regime to the thermal hopping
regime is first-order. We have argued that for large values of $f$ the
transition is second order. We have also calculated the action
analytically at zero temperature by assuming an appropriate ansatz
solution for the bounce. The result is in good agreement with the
exact numerical result in the thin-wall approximation (TWA). 

It is convenient to use the following form of the potential
proposed by Adams \cite{adams} 
\begin{equation}
 U(\varphi) = {1 \over 4}\varphi^4 - \varphi^3 + {\d \over 2} 
\varphi^2 ~~ , \label{adam:1}
\end{equation}
where $ 0 \le \delta \le 2$. Any quartic potential can be reduced to
this form by shifting and rescaling of the field. In this paper we
extend our calculation of the action to finite temperatures and study
the nature of the transition. We propose a general ansatz at finite
temperature in the thin wall ($\d \to 2$) and thick wall ($\d \to 0$)
limits. We find that for a thin wall the transition is first-order
while for thick wall it is second-order. This result is in agreement
with Ferrera\cite{ferrera} who found that only for very large wall
thickness (i.e., $\d \sim 0.6$) a second order transition takes place,
while for all other cases a first order transition occurs. We would
like to point out that the analytical approach is very much easier
than the numerical one and less time consuming \cite{hatem}.

 In Sec. II we present our analytic calculation for the action
at zero temperature and high temperature, including some earlier work
which is presented for completeness. In Sec. III, we extend the
calculations to finite temperatures and discuss the nature of the phase
transition. Section IV contains our results for intermediate wall
sizes. Section V contains our conclusions. The algebraic expressions
for the integrals appearing in the analytic formalism are given in the
appendices $A$ and $B$.

\end{section}
%
%
\begin{section}*{II. ANALYTICAL SOLUTION FOR ZERO AND HIGH TEMPERATURES}
 We use the following potential (see \cite{adams} and \cite{hatem}) to
calculate the action analytically in two extreme limits: the thin-wall
and thick-wall limits
\be
U(\varphi) = {1 \over 4}\varphi^4 - \varphi^3 + 
{\d \over 2} \varphi^2 ~ .
\ee
%
\vskip 1.2cm
\noindent
\underline{\bf A. { Thin-wall limit at zero temperature: $\d \to 2$}}

We find that an analytic solution for the bounce of the form of a Fermi
function:
\be
\varphi =  {\g \over \s}  ~ , \label{twa1:1} 
\ee
where  $\rho=\sqrt{\vec x^2+\tau^2}$, $R$ is the radius of the bubble
and $\l$ its width, acts like a bounce in the TWA and leads to the
correct value for $S_4$, the action at zero temperature.

Here the false minimum of the potential is at $\varphi=0$ and the true
minimum  lies between 2 (for $\d=2$) and 3 (for $\d=0$). The parameter
$\g$ is approximately equal to the true minimum in the thin wall
approximation. The bounce has values $\varphi = \g$ at $\rho=0$ 
and $0$ at $\rho \to \infty$. These boundary conditions 
are satisfied by Eq.~(\ref{twa1:1}).

To evaluate $\g$, $R$, and $\l$, we substitute the ansatz
(\ref{twa1:1}) in the equation of motion : 
\be 
{d^2\varphi \over d\rho^2}+ {3 \over \rho}{d\varphi \over d\rho} = 
\varphi^3 - 3 \varphi^2 + \d \varphi ~. \label{twa2:2}
\ee
Then the left-hand side (L.H.S.) and the right-hand side (R.H.S.) are
respectively
\bea
L.H.S.= {{8 \g \rho^2 / \l^4} \over (\s)^3} & + &{{\g (-12 \rho^2 / \l^4 
+ 8 / \l^2)} \over (\s)^2 }  \nonumber \\ [0.3cm]
 & + & {{\g (4 \rho^2 / \l^4 - 8 / \l^2)} \over \s }
\label{twa3:3} ~. 
\eea
\bea
R.H.S.= {{\g \d} \over \s}
 & - & {{ 3 \g^2} \over (\s)^2 } \nonumber \\ [0.3cm]
 & + & {{\g^3} \over (\s)^3}
\label{twa4:4} ~.
\eea
 In the TWA, the solution is constant except in a narrow region near the 
wall at $\rho=R$. So, we replace in Eq.~(\ref{twa3:3}) 
\bea
\mathrm{
8 \rho^2 / \l^4 \quad by \quad {{ 8 R^2} \over \l^4} ( 1- {\it a} \l^2 /R^2) 
\quad in \quad the \quad { 1 \over (\s)^3} \quad term }
\label{twa5:5} ~, \\ [0.3cm]
\mathrm{
8 / \l^2 -12 \rho^2 / \l^4 \quad by \quad {-}{{ 12 R^2} \over \l^4}
(1- {\it b} \l^2 / R^2 ) \quad in \quad the \quad { 1 \over (\s)^2} \quad 
\quad term } ~, \label{twa6:6} \\ [0.3cm]
\mathrm{
4 \rho^2 / \l^4- 8 / \l^2 \quad by \quad {{4 R^2} \over \l^4}
(1- {\it d}  \l^2 / R^2) \quad in \quad the \quad { 1 \over \s} \quad term
}
\label{twa7:7} ~ ,
\eea
where $a$, $b$ and $d$ are parameters to be determined later.

Comparing Eq.~(\ref{twa3:3}) with Eq.~(\ref{twa4:4}) in the
range
$ R^2(1- \l^2 / R^2)=R^2-\l^2 < \rho^2 < R^2 + \l^2 =R^2(1+ \l^2 / R^2)$ 
where $\rho^2 \simeq R^2$  as $\l^2 / R^2 << 1$ , we have :
\bea
{\g^2 \over 8} = {{ R^2} \over \l^4} (1- a \l^2 / R^2) ~ , 
\nonumber \\ [0.3cm]
{\g \over 4} = {{R^2} \over \l^4}( 1-b \l^2 / R^2) ~ ,
\label{twa10:10} \\ [0.3cm]
{ \d \over 4} = {{R^2 \over \l^4} ( 1- d \l^2 / R^2)} ~ ,
\nonumber 
\eea
We can now evaluate the zero-temperature action $S_4$ : 
\be
S_4= 2 \pi^2 \> \int_0^\infty d\rho \> \rho^3 \left[{1 \over 2} 
\Bigg({d\varphi \over d\rho}\Bigg)^2 + U(\varphi) \right] ~ .
\label{twa12:12} 
\ee
Substituting Eq.~(\ref{twa1:1}) in Eq.~(\ref{twa12:12}) and
integrating we get
\bea
S_4 & = & 2 \pi^2 \g^2 R^4 \Bigg[ {1 \over {6 \l^2}}
\Bigg(1+\Big({\pi^2 \over 3}-2 \Big) { \l^4 \over R^4} \Bigg) + 
{\d \over 8} \Bigg(1- 2 \l^2 / R^2 +{ \pi^2 \over 3}{ \l^4 \over
R^4}\Bigg)  \nonumber \\ 
& & + {\g \over 4} \Bigg(1- 3 \l^2 / R^2 + \Big({\pi^2 \over 3} +
1\Big) \l^4/R^4 \Bigg) 
\nonumber \\
& & + { \g^2  \over 16} \Bigg (1- {{11 \l^2} \over {3 R^2}}+
\Big({\pi^2 \over 3} + 2 \Big) \l^4/R^4 \Bigg) \Bigg] ~ .
\label{twa13:13} 
\eea
We now determine the parameters $a$, $b$, and $d$ by demanding 
${dS_4 / dR^2}={dS_4 / d\l^2}={dS_4 / d\g}=0$.
 Differentiating Eq.~(\ref{twa13:13}) and using
Eq.~(\ref{twa10:10}), we find that, to leading order in $\l^2/R^2$,
\bea
4b- 2a -2d +1 =0 ~ , \nonumber \\ [0.3cm]
3b -2a - d =0 ~ ,  \\ [0.3cm] \label{twa14:14}
3b- 11a /6 -d =0 ~ , \nonumber
\eea
which leads to $a=0$, $b=1/2$ and $d=3/2$. Using Eq.~(\ref{twa10:10}),
we can rewrite Eq.~(\ref{twa13:13}) as :
\bea 
S_4 = {2 \pi^2}{ 8 R^6 \over \l^6}\Bigg[ \Big(1/3-d/2-a/2+b\Big) & + & 
{\l^2 \over R^2} \Big(d - 3b + 11 a /6\Big) \Bigg] ~ , \label{twa15:15}
\eea
where the coefficient of $\l^4 \over R^4$ evaluated by the usual methods 
of statistical mechanics for the Fermi function vanishes. This gives
\bea
S_4 = {4 \pi^2 \over 3} { R^6 \over \l^6} +O({\l^6 \over R^6}) ~ . 
\label{twa16:16}
\eea
The quantities $\g$, $R$ and $\l$ are determined from
Eq.~(\ref{twa10:10}) using the values of $a$, $b$, and $d$. So we have 
\be
\g^2 - 2 \g {{d-a \over {d-b}}} + 2 \d {{b-a \over d-b}} = 0 ~ ,
\label{twa25:25}\\ [0.3cm]
\ee
which gives
\be
\l^2 = {8(b-a) \over {\g^2-2\g}}={{4 (d-b) \over {\g -\d}}} ~ , 
\label{twa17:17}
\ee
with $\g$ given by Eq.~(\ref{twa25:25}). We have then, for
$\d=1.9$, $\g=2.1$, which implies that 
$R^2/\l^2={\g^2b}/({\g-\d})=11$. Thus we have  
\be
 S_4 = { 4 \pi^2 \over 3} (11)^3 ~ , \label{twa18:18}
\ee
while the action from the TWA formula is ( see \cite{adams}) 
$S_{TW} = { 4 \pi^2 \over 3} (10)^3$ for $\d=1.9$.  The departure from
TWA, ${S_4 / S_{TW}}=1.33$, is in agreement with 
Ref.~\cite{adams}. The expressions seem certainly valid 
for values of $\d$ in the range 2.0 to 1.8 .

\vskip 1.2cm
\noindent
\underline{\bf B. { Thin-wall limit at high temperature: $\d \to 2$}}

The bounce takes the following form
\be
\varphi =  {\g \over \ss}  ~ , \label{twa03} 
\ee
where $r^2=\vec x^2$ and the other parameters $R$ and $\l$ have the same
physical significance in three dimensions. The boundary conditions
are $\varphi = \g$ at $r=0$ and $\varphi=0$ at $r \to \infty$,
$d\vp/dr=0$ at $r=0$. The equation of motion is
\be 
{d^2\varphi \over d r^2}+ {2 \over r}{d\varphi \over d r} = 
\varphi^3 - 3 \varphi^2 + \d \varphi ~. \label{eomo3}
\ee
As in the earlier subsection, we substitute the ansatz bounce in the
equation of motion and assume the solution is constant except in
a narrow region near the wall at $r=R$. The resulting equations have a
structure similar to that of Eqs. (\ref{twa3:3}) to (\ref{twa10:10}).
\be
S_3= 4 \pi \> \int_0^\infty d r \> r^2 \left[{1 \over 2} 
\Bigg({d\varphi \over d r}\Bigg)^2 + U(\varphi) \right] ~ . 
\ee
After substituting the bounce Eq.~(\ref{twa03}) in the action and
integrating, we get the following 
\bea
S_3 & = & \frac{4 \pi}{3} \g^2 R^3 \Bigg[\frac{1}{2 \l^2}
\Bigg(1+\Big(\frac{\pi^2}{8}-\frac{3}{4}\Big)\frac{\l^4}{R^4}\Bigg) 
\nonumber \\ [0.3cm]
& + &
\frac{\g^2}{4}
\Bigg(1-\frac{11}{4}\frac{\l^2}{R^2}+\Big(\frac{\pi^2}{8}+\frac{3}
{4}\Big)\frac{\l^4}{R^4}\Bigg)
\nonumber \\ [0.3cm]
& - &
\g \Bigg(1-\frac{9}{4}\frac{\l^2}{R^2} +
\Big(\frac{\pi^2}{8}+\frac{3}{8} \Big) \frac{\l^4}{R^4} \Bigg)
\nonumber \\ [0.3cm]
& + &
\frac{\d}{2} \Bigg(1-\frac{3}{2} \frac{\l^2}{R^2} +
\frac{\pi^2}{8}\frac{\l^4}{R^4} \Bigg) \Bigg] ~.
\eea
In terms of the parameters $a$, $b$ and $d$, the action takes the
simpler form 
\be
S_3= \frac{32 \pi}{3}\frac{R^5}{\l^4} \Bigg(1-2a+4b-2d +
\frac{\l^2}{R^2} \Big(\frac{11}{2}a -9b +3d \Big) \Bigg) 
\label{actiontwa03} ~,
\ee 
where the relations between $a$, $b$ and $d$ to leading order in
$\l^2/R^2$ are
\bea
-2a + 4b -2d +\frac{2}{3} = 0 ~, \nonumber \\ [0.3cm]
-\frac{11}{2} a + 9b - 3 d = 0 ~, \nonumber \\
[0.3cm] 
-2a + 3b -d = 0 ~,
\eea
this leads to $a=0$, $b=1/3$ and $d=1$.
\vskip 1cm
Hence the action in Eq.~(\ref{actiontwa03}) is reduced to
\be
S_3 = \frac{32\pi}{9} \frac{R^5}{\l^6} + O(\frac{\l^6}{R^6}) ~,
\ee
which agrees with the TWA formula and the expression of Adams
\cite{adams}.

%
\vskip 1.2 cm
\noindent
\underline{\bf C. { Thick-wall limit at zero temperature: $\d  \to 0$}}
 
 The form of the bounce in Eq.~(\ref{twa1:1}) suggests that the thick
wall limit, which would correspond to small values of $R^2/\l^2$, would
be obtained by approximating the Fermi function by the 
Maxwell-Boltzmann function, which leads to a Gaussian:
\be
 \varphi = \g e^{-\rho^2/\l^2} ~ . \label{twa19:19}
\ee
 The action for this form of bounce is found to be
\be
 S_4 = { {\pi^2 \g^2 \l^4}} \Bigg[ {1 \over {2\l^2}} 
+ { \d \over
8} - {\g \over 9} + {\g^2 \over 64}\Bigg] ~. \label{twa20:20}
\ee
For small values of $R^2/\l^2$, the relation between the parameters in
the bounce Eq.~(\ref{twa19:19}) and the constants $a$, $b$ and $d$ is
given by 
\begin{equation}
{\g^2 \over 8} = -{a \over \l^2} ~~, \quad 
{\g \over 4} = -{ b \over \l^2} ~~, \quad 
{ \d \over 4} = - {d \over\l^2} ~.
\end{equation}
Note that in this case $\g \ll 1$, so $\g^2$ is negligible.

The values of $b$ and $d$ are again obtained by demanding
$dS_4/d\l^2=dS_4/d\g = 0$. This gives $b=-9/8$, $d=-1/2$, giving
\be
\l^2 = {2 \over \d} ~, \quad \g={9 \over 4} \d ~ .
\ee
This yields the action
\be
 S_4 = {{4 \pi^2} \over 3} (1.9) \d \Big(1+ O({R^2 \over \l^2})
\Big) ~ . \label{dwt1:1}
\ee
 The ratio of the action to the TWA value is 
\be
 R_4 = { S_4 \over S_{TW}} = 1.9 \d (2-\d)^3 ~ .
\ee
For $\d=0.1$, $R_4=1.31$, which agrees with Adams' result.

We find the ratio of the actions when we take the next order in $\d$
in $S_4$ to be
\be
R_4 \propto \d \Big( 1 - 0.826 \> \d -0.150 \> \d^2 + 0.320 \> \d^3
\Big) ~,
\ee
as compared to Adams'
\be
R_4 \propto  \d \Big( 1-0.8 \> \d + 0.15 \> \d^2 \Big).
\ee
We find the two expressions agree for $0<\d<0.5$.

%
\vskip 1.2 cm
\noindent
\underline{\bf D. { Thick-wall limit at high temperature: $\d  \to 0$}}

At higher temperatures the bounce takes the form
\be
 \varphi = \g e^{-r^2/\l^2} ~,
\ee
with the action

\be
S_3 = 4 \pi^{3/2} \g^2 \l^3 \Bigg[
\frac{3}{16 \sqrt{2}}\frac{1}{\l^2} + \frac{\g^2}{128} - \frac{\g}{12
\sqrt{3}} + \frac{\d}{16 \sqrt{2}} \Bigg] ~. \label{actionthick03}
\ee
Defining $\g/4=-b/\l^2$, $\d/4=-d/\l^2$ and neglecting $\g^2$, we find
$b$ and $d$ as before by demanding $dS_3/d\l=dS_3/d\g=0$. The relation
between $b$ and $d$ is given by 
\bea
3 + 12 (\frac{2}{3})^{3/2} b - 4 d = 0 ~,
\nonumber \\ [0.3cm]
\frac{3}{4} +6 (\frac{2}{3})^{3/2} b -3 d =0 ~,
\eea
which leads to $b=-\frac{3}{4}(3/2)^{1/2}$ and $d=-3/4$, giving
$\l^2=3/\d$ and $\g={\sqrt{3/2}} \d$. The action can be simplified to
\be
S_3 = \frac{3 {\sqrt{3}}}{4\sqrt{2}} \pi^{3/2} \d^{3/2}
\Big(1+\frac{9{\sqrt{2}}}{32} \d \Big) ~.
\ee
Following Adams \cite{adams}, we can calculate the ratio of this 
action to the thick-wall action and we find 
\be
R_3=5.828 \d^{3/2} \Big(1-0.602 \> \d - 0.148 \> \d^2 + 0.099
\> \d^3 \Big) ~,
\ee
and this matches very well with Adams' result
\be
R_3=5.864 \d^{3/2} \Big(1-0.667 \> \d +
0.099 \> \d^2 \Big) ~.
\ee
We find both compare well for values of $\d$ from $0$ to $0.5$.

Thus, the form of the bounce given by Eqs.~(\ref{twa1:1}) and
(\ref{twa03}) seems valid over the whole range of $\d$ (from 0 to 2),
and in the two extreme limits is amenable to analytic
calculations. This suggests that we look for an interpolating form
valid at all temperatures that reduces to Eq.~(\ref{twa1:1}) at $T=0$
and to Eq.~(\ref{twa03}) at high temperatures.

\end{section}

%
%
\begin {section}*{III. RESULTS FOR INTERMEDIATE TEMPERATURES}

\underline{\bf A. { Thin-wall limit : $\d \to 2$}}

The action at finite temperature of a single scalar field $\vp$ is
given by the following formula
\be
S(T)= 4\pi \int_{-\b/2}^{\b/2} d\tau \int_0^{\infty} dr r^2
\Bigg[\frac{1}{2} \Bigg(\frac{\partial \vp}{\partial\tau}\Bigg)^2 + 
\frac{1}{2} \Bigg(\frac{\partial \vp}{\partial r}\Bigg)^2 + U(\vp)
\Bigg] ~. \label{action}
\ee
The equation of motion derived from the above action is given by the
following expression
\be
{\partial^2\vp \over \partial\tau^2} +
{\partial^2\vp \over \partial r^2} + 
{2 \over r }{\partial\vp \over \partial r} = 
{\partial U(\vp,T) \over \partial\vp} ~, \label{eom}
\ee
with boundary conditions 
\be
\vp \to \vp_- \quad as \quad r \to \infty ,\quad 
\partial\vp/\partial\tau = 0 \quad at \quad \tau = \pm \beta/2,0
~, \label{bcs} 
\ee
where $\vp_-$ is the false vacuum of the potential $U$, $\beta$ is
the period of the solution and $r=\sqrt{\vec{x}^2}$

We assume for the solution of the equation of motion the following
ansatz 
\be
\vp(r,\tau)=\frac{\g}{\t} ~, \label{ansatz}
\ee
which is periodic in the interval $(-\b/2,\b/2)$ and satisfies the
required boundary conditions (Eq.~(\ref{bcs})), viz
\be
\frac{\partial\vp}{\partial r}=0 ~\mathrm{at}~ r=0,
~\frac{\partial\vp}{\partial\tau}=0 
~\mathrm{at}~ \tau=0 ~\mathrm{and} ~ \pm\b/2, ~\mathrm{and} ~\vp=0
~\mathrm{as}~ r \to \infty ~.
\ee
Note that for the potential given by Eq.~(\ref{adam:1}), $\vp=0$ is
always the false vacuum.

We evaluate the action for potential given by Eq.~(\ref{adam:1}) 
\be
U(\vp)=\frac{1}{4} \vp^4 -\vp^3 +\frac{\d}{2}\vp^2~, 0 \le\d\le 2
~. \label{pot}
\ee
After substituting the ansatz function
Eq.~(\ref{ansatz}) into the equation of motion Eq.~(\ref{eom}), we have
\bea
\frac{\g}{(\t)^3} \Bigg[ \frac{8 r^2}{\l^4}+\frac{2\b^2}{\pi^2 \l^4}
\sin^2(\arg) \Bigg] 
\nonumber \\  [0.3cm] 
+ \frac{\g}{(\t)^2} 
\Bigg[ -\frac{12
r^2}{\l^4}+\frac{6}{\l^2}+\frac{2}{\l^2} \cos(\arg) -
\frac{3\b^2}{\pi^2 \l^4} \sin^2(\arg) \Bigg]  
\nonumber \\  [0.3cm]
+ \frac{\g}{\t} 
\Bigg[ \frac{4 r^2}{\l^4} - \frac{6}{\l^2} + 
\frac{\b^2}{\pi^2 \l^4} \sin^2(\arg) - \frac{2}{\l^2} \cos(\arg)
\Bigg] 
\nonumber \\  [0.3cm]
= \frac{\g^3}{(\t)^3} -\frac{3\g^3}{(\t)^2} 
\nonumber \\  [0.3cm]
+ \frac{\g \d}{\t} ~.
\eea
Equating terms with different powers of exponentials separately, we
have with $r^2+\frac{\b^2}{4\pi^2} \sin^2(\arg)\approx R^2$
\bea
\frac{\g^2}{8}=\frac{R^2}{\l^4} \Bigg[ 1-a\frac{\l^2}{R^2} \Bigg] ~.
\nonumber \\ [0.3cm]
\frac{\g}{4}=\frac{R^2}{\l^4} \Bigg[ 1 - b \frac{\l^2}{R^2} \Bigg] ~.
\nonumber \\ [0.3cm] 
\frac{\d}{4}= \frac{R^2}{\l^4} \Bigg[ 1 -d \frac{\l^2}{R^2} \Bigg] ~.
\label{abd} 
\eea
As in the last section, the parameters $a$, $b$ and $d$ are found by
the requirement that the variation of $S(T)$ with respect to the
parameters $R$, $\l$ and $\g$ in Eq.~(\ref{ansatz}) vanish.

The integrals in the action are obtained in powers of $\l^2 /R^2$
using the usual methods for evaluating integrals of the Fermi function
(see e.g. Huang \cite{huang}). We 
get 
\bea
S(T) & = & \frac{8\pi}{3} R^4 \ka E_3 \, \g^2 \Bigg[ \frac{1}{2\l^2}
\Bigg( 1 + \ka^2 \frac{E_T}{E_3} +
\Big(\frac{\pi^2}{8}-\frac{3}{4}\Big)  
\frac{E_0}{E_3} \frac{\l^4}{R^4} \Bigg) 
\nonumber \\ [0.3cm]
& + &
\frac{\g^2}{4} \Bigg(1 - \frac{11}{4} \frac{E_1}{E_3} \frac{\l^2}{R^2}
+ 
\Big(\frac{\pi^2}{8}+\frac{3}{4}\Big) \frac{E_0}{E_3}
\frac{\l^4}{R^4}\Bigg)   
\nonumber \\ [0.3cm]
& - &
\g \Bigg(1 - \frac{9}{4} \frac{E_1}{E_3} \frac{\l^2}{R^2} +
\Big(\frac{\pi^2}{8}+\frac{3}{8}\Big)
\frac{E_0}{E_3}\frac{\l^4}{R^4}\Bigg)   
\nonumber \\ [0.3cm]
& + & \frac{\d}{2} \Bigg(1 - \frac{2}{3} \frac{E_1}{E_3}
\frac{\l^2}{R^2}+ 
\frac{\pi^2}{8} \frac{E_0}{E_3} \frac{\l^4}{R^4} \Bigg) \Bigg] ~,
\label{actionint} 
\eea 
where
\bea
E_0 & = & \int_0^1 \frac{dt}{\sqrt{1-t^2}\sqrt{1-\ka^2 t^2}} ~,
\nonumber \\ [0.3cm]
E_1 & = & \int_0^1 \frac{dt\sqrt{1-\ka^2 t^2}}{\sqrt{1-t^2}} ~,
\nonumber \\ [0.3cm]
E_3 & = & \int_0^1 \frac{dt(1-\ka^2 t^2)^{3/2}}{\sqrt{1-t^2}} ~,
\nonumber \\ [0.3cm]
E_T & = & \int_0^1 \frac{dt (1-t^2) t^2 \sqrt{1-\ka^2 
t^2}}{\sqrt{1-t^2}} \nonumber  ~,
\eea
are complete elliptic integrals which can be represented in
terms of the basic complete elliptic integrals $E_0$ and
$E_1$ (see Appendix \ref{app1}), $\ka=\frac{\b}{\pi R}$
and $t=\sin\frac{\pi}{\b}\tau$ 

We now determine the parameters $a$, $b$ and $d$ by demanding the
vanishing of $dS(T)/dR^2$, $dS(T)/d\l^2$ and $dS(T)/d\g$. Differentiating
Eq.~(\ref{actionint}) and using Eq.~(\ref{abd}), we find that to
leading order in $\l^2/R^2$,
\bea
-2a+4b-2d+\frac{1}{2} +\frac{1}{2} \frac{E_1-2\ka^2 E_1'}{3E_3-2\ka^2
E_3'} + \frac{\ka^2}{2} \frac{E_T -2 \ka^2 E_T'}{3E_3-2\ka^2 E_3'} = 0
~.
\nonumber \\ [0.3cm]
a \Bigg(11 g - 2 \g \epsilon_T \frac{E_3}{E_1} \Bigg) + 
b \Bigg( \g^2 \epsilon_T \frac{E_3}{E_1} -18 g \Bigg) +
6 g d = 0 ~. 
\nonumber \\ [0.3cm]
-2a+3b-d-\frac{1}{4} \epsilon_T = 0 ~, \label{derivative}
\eea
where
$g=\g^2-2\g$, $E_1'$ is the derivative of $E_1$ with respect to
$\ka^2$ (and similarly for $E_T'$ and $E_3'$), and
$\epsilon_T=E_1/E_3-\ka^2 E_T/E_3 -1$.

By using Eq.~(\ref{derivative}), we can find a relation between the
constants $a$, $b$ and $d$,
\bea
d-a=\frac{3}{4}(1+c)+\frac{\epsilon_T}{2} ~,
\nonumber \\ [0.3cm]
d-b=\frac{1}{2}(1+c)+\frac{\epsilon_T}{4} ~,
\nonumber \\ [0.3cm]
b-a=\frac{1}{4}(1+c+\epsilon_T) ~,
\nonumber \\
a=\frac{\epsilon_T [\g^2(b-a)E_3/(E_1g)-1.5]}{(1- \epsilon_T
E_3/E_1)}~, 
\eea
where $c$ is given by the following expression
\be
c=\frac{E_1-2\ka^2 E_1'}{3E_3-2\ka^2 E_3'} +
\frac{\ka^2(E_T-2\ka^2E_T')}{3E_3-2\ka^2E_3'} ~. \nonumber
\ee
Thus Eq.~(\ref{actionint}) can be expressed in terms of $a$, $b$ and
$d$. It reads as
\be
S(T)=\frac{64}{3} \pi \ka \Bigg(\frac{R}{\l}\Bigg)^6
\Bigg[\Bigg(1-a\frac{\l^2}{R^2} \Bigg)
\Bigg(1-2(b-a)\frac{E_3}{E_1}\Bigg)
-\frac{\l^2}{R^2}\Bigg(\frac{3}{4}\epsilon_T + \frac{a}{2}\Bigg)
\Bigg]  ~.
\label{actionf}
\ee
With these expressions we calculate the values of $a$, $b$ and $d$
and also for the action $S(T)$. In calculating $S(T)$ for $\b \to
\infty$, the integrals $E_{i4}$ are to be used (see Appendix
\ref{app1}). In these cases $\ka >1$, and we restrict the upper limit
of the elliptic integrals to $1/\ka$ as they become complex for larger
values of $\ka$.

One observation is the occurrence of a singularity in $(b-a)$ (which
is directly proportional to $R$) due to $E_0(\ka)$ becoming singular
at $\ka=1$ (i.e. at $\beta=\pi R$). The values of $S(T)$ are obtained
and plotted for $\d=1.9$ and $1.85$ as shown in Figs. 1 and 2
respectively. The temperature $T_\star$ ($\beta_\star=1/T_\star$) is
defined by \cite{hatem} $\beta_\star=S_4/S_3$. The transition point
$\beta_c$ can in principle be different from $\beta_\star$, but in the
TWA they are equal \cite{hatem}. In our result (Figs. 1 and 2) we can
readily determine $\beta_\star$ by extending the horizontal part of
the curve to the left. For Fig. 2, for example, this yields $\beta_\star
\approx 25$, which is close to the value of $\beta_c$ obtained
numerically as well as analytically \cite{hatem}. We conclude,
therefore, that the singularity at $\beta \approx 45$ in Fig. 2 is an
artifact of the method, and does not represent the transition
point. The phase transition actually takes place at a much lower value
of $\beta_c$, and is first-order. We suggest that the same
prescription be 
used to determine the transition point for other values of $\delta$
within the TWA. This means that the phase transition is first order, 
which is consistent with the definitions available in the literature
(see, for example, \cite{huang}). 

 It can be shown that in the limit of zero temperature (i.e. $\kappa
\to \infty$), the action in Eq.~(\ref{actionf}) reduces to the
action given by Eq.~(\ref{twa15:15}), while at high temperature
(i.e. $\kappa \to 0$), it reduces to the one given by
Eq.~(\ref{actiontwa03}). 

%
\vskip 1.2cm
\noindent
\underline{\bf B. { Thick-wall limit: $\d  \to 0$}}

The form of the bounce in Eq.~(\ref{ansatz}) suggests that the thick
wall limit, which would correspond to small values of $R^2/\l^2$,
would be obtained by approximating the Fermi function by the
Maxwell-Boltzmann function, which leads to a Gaussian:
\be
\vp(r,\tau)= \g
e^{-(r^2+\frac{\b^2}{\pi^2}\sin^2(\frac{\pi\tau}{\beta}))/\l^2} ~,
\label{ansatz1} 
\ee
which satisfies the boundary conditions given by Eq.~(\ref{bcs}).

The action for this form of bounce is found to be
\bea
S(T) & = &2\pi^2 \g^2 \Gamma(3/2) \l x (\frac{\l^2}{2})^{3/2} e^{-x^2}
I_0(x^2) 
\Bigg[\frac{3}{2\l^2} \Bigg(1+\frac{1}{3}\frac{I_1(x^2)}{I_0(x^2)}\Bigg) +
\frac{1}{4} (\frac{1}{2})^{3/2} \g^2 e^{-x^2} \frac{I_0(2
x^2)}{I_0(x^2)} \nonumber \\ [0.3cm]
&- & (\frac{2}{3})^{3/2} \g e^{-x^2/2}
\frac{I_0(\frac{3}{2}x^2)}{I_0(x^2)} 
+\frac{\d}{2} \Bigg] ~, \label{thickaction}
\eea
where $x=\frac{\beta}{\pi \l}$ and $I_\nu(x^2)$ are the modified Bessel
functions.

Eq.~(\ref{abd}) then reduces to
\be
\frac{\g^2}{8}=-\frac{a}{\l^2}, ~~ \frac{\g}{4}=-\frac{b}{\l^2},~~
\frac{\d}{4}=-\frac{d}{\l^2} ~.
\ee
Here again as in Sec. II, $\g \ll 1$, so $\g^2$ is negligible.

The values of $b$ and $d$ are again obtained by demanding $dS(T)/d\g=
dS(T)/d\l=0$. This gives the following
\be
3\Bigg(1+\frac{1}{3} \frac{I_1(x^2)}{I_0(x^2)} \Bigg)+12 
(\frac{2}{3})^{3/2} b e^{-x^2/2}
\frac{I_0({\frac{3}{2}}x^2)}{I_0(x^2)} 
-4 d =0 ~. \label{gammder}
\ee
\bea
\frac{3}{4} e^{-x^2} I_0(x^2)+ x^2 e^{-x^2}
\Bigg(I_0(x^2)-I_1(x^2)\Bigg)
+\frac{3}{4} e^{-x^2} I_1(x^2)+ \nonumber \\ 
3(\frac{2}{3})^{3/2} F b e^{-\frac{3}{2}x^{3/2}}I_0(\frac{3}{2}x^2) 
-d E e^{-x^2} I_0(x^2) =0 ~, \label{lamder}
\eea
where
\bea
E=3 + 2 x^2 \Bigg(1-\frac{I_1(x^2)}{I_0(x^2)} \Bigg) ~, \nonumber \\
[0.3cm] 
F= 2 + 2 x^2
\Bigg(1-\frac{I_1(\frac{3}{2}x^2)}{I_0(\frac{3}{2}x^2)}\Bigg) ~.  
\eea
Using Eqs.~(\ref{gammder}) and (\ref{lamder}), the values of $b$ and
$d$ are given by 
\be
b=\frac{\frac{E}{4} \Big(3 I_0(x^2)+I_1(x^2) \Big) - \frac{3}{4} e^{-x^2}
\Big(I_1(x^2)+I_0(x^2) \Big)- x^2 e^{-x^2} \Big(I_0(x^2)-I_1(x^2) \Big)}
{3(\frac{2}{3})^{3/2} (F-E) e^{-\frac{3}{2}x^2} I_0(\frac{3}{2}x^2)} ~.
\ee
\be
d=\frac{\frac{F}{4} \Big(3 I_0(x^2)+I_1(x^2) \Big) - \frac{3}{4} e^{-x^2}
\Big(I_1(x^2)+I_0(x^2) \Big)- x^2 e^{-x^2} \Big(I_0(x^2)-I_1(x^2) \Big)}
{(F-E) e^{-x^2} I_0(x^2)} ~.
\ee
This yields the action
\be
S(T)=32 \pi^2 (\frac{1}{2})^{3/2} (\frac{b}{\l^2})^2 \Gamma(3/2) \l^2 x 
e^{-x^2} I_0(x^2) 
\Bigg[\frac{3}{2} \Bigg(1+\frac{1}{3}\frac{I_1(x^2)}{I_0(x^2)}\Bigg) +
4 (\frac{2}{3})^{3/2} b e^{-x^2/2}
\frac{I_0(\frac{3}{2}x^2)}{I_0(x^2)} 
-2d \Bigg] ~. 
\ee
For a given value of temperature (i.e. $x$) we can calculate $b$ and
$d$. Hence $\g$ and $\l$ are determined. Thus we can calculate the
action at different values of temperatures. Figures 3 and 4 
show the value of the action at different values of inverse of
temperatures for $\d=0.3$ and $\d=0.1$ respectively. As we can see
from the figure, the action goes smoothly 
from the zero temperature regime to high temperature regime without
any singularity at the transition point. This means that in the
thick-wall limit the transition is second order.

 It can be show easily that in the limit of zero temperature (i.e. $x
\to \infty$), the action in Eq.~(\ref{thickaction}) reduces to that in
Eq.~(\ref{twa20:20}). Also we can recover the values of $b$ and $d$,
i.e., $b=-9/8$ and $d=-1/2$. In the limit of high temperatures
(i.e. $x \to 0$), Eq.~(\ref{thickaction}) reduces to
Eq.~(\ref{actionthick03}) and the values of $b$ and $d$ are recovered.

\end{section}

%
\begin {section}*{IV. RESULTS FOR INTERMEDIATE WALL SIZES}

The results in Sec. II and III apply to situations where the bubble
size $R$ is large (TWA) or the bubble has no radius but only a
(exponentially) decreasing wall. In this section we consider
expressions for the action with $y_\l=R^2/\l^2$ a small quantity.

\vskip 1.2cm
\noindent
\underline{\bf A. zero temperature}

For the zero temperature or O(4) invariant bounce, an exact expression
for the action can be found and it can be approximated for large or
small values of $y_\l$. For small $y_\l$ we have
\bea
S_4 &=& 2\pi^2 \g^2 \l^4 \Bigg[\frac{1}{\l^2} \Bigg(\frac{\pi^2}{36}
-\frac{1}{6}+y_\l \Big(\frac{\ln{2}}{3}-\frac{1}{12}\Big)
\Bigg) 
+\frac{\d}{4} \Bigg(\frac{\pi^2}{12}-\ln{2}+y_\l
\Big(\ln{2}-\frac{1}{2}\Big)\Bigg) \nonumber \\ [0.3cm]
& - & \frac{\g}{2}\Bigg(\frac{\pi^2}{12}+\frac{1}{4}-\frac{3}{2} \ln{2}
+ y_\l \Big(\ln{2} -\frac{5}{8}\Big)\Bigg)  \nonumber \\
[0.3cm] 
& + & \frac{\g^2}{8}\Bigg(\frac{\pi^2}{12}+\frac{11}{24}-
\frac{11}{6}\ln{2} + y_\l \Big(\ln{2} -\frac{2}{3} \Big)
\Bigg) \Bigg] ~.
\eea
Extremization with respect to $\g$ and $\l$ leads to $a=-14.9924$,
$b=-6.2335$ and $d=-1.5785$ with values of $\g$ given by the equation
\be
\g^2-5.76 \> \g + 3.76 \> d=0, ~~ 3.7<\g<5.8, ~~ 0<\d<2 ~.
\ee
The value of $y_\l$ for $\d=0$ is $y_\l \approx -1.5$ though the
expression for the action is not valid as now $|y_\l| >
1$. Interestingly, $3 e^{-y_\l}$ corresponds to the value of $\g
(\approx \d)$ obtained in Sec. II. As we expect the value of $\g$ to
be $3$ (value of $\vp$ at the true minimum) it appears that the
limiting solution is of the form
\be
\vp = 3 \> e^{-y_\l} e^{-\rho^2/\l^2} ~.
\ee
We see that $y-{\l} \approx 0$ for $\d \approx 1.2$. The action
calculated agrees with the action calculated numerically by Adams
\cite{adams} in the region $1.2<\d<1.4$.

%
\vskip 1.2cm
\noindent
\underline{\bf B. High temperature}

In this case we start with the expression for small $y_\l$ given by
\bea
S_3 & = & 4 \pi^{3/2} \g^2 \l^3
\Bigg[\frac{1}{\l^2}\frac{3}{16\sqrt{2}} 
\Bigg(1+ y_\l \Big(2-\frac{16\sqrt{2}}{9\sqrt{3}} \Big) \Bigg)
+\frac{\d}{16\sqrt{2}} \Bigg( 1+y_\l \Big(2-\frac{4\sqrt{2}}{3\sqrt{3}}
\Big) \Bigg) \nonumber \\ [0.3cm]
& - &\frac{\g}{12\sqrt{3}} \Bigg( 1+y_\l \Big(3-\frac{9\sqrt{3}}{8}
\Big) \Bigg)
+\frac{\g^2}{128} \Bigg( 1+y_\l \Big(4-\frac{32\sqrt{5}}{25}
\Big) \Bigg) \Bigg] ~.
\eea
Minimization leads to $d=-1.588$, $b=-2.403$, and $a=-2.2857$. Again
we find the values of $y_\l \approx 0$ for $\d=1.25$ and the
expressions are valid between $1.1<\d<1.4$. Thus we have (semi)
analytic expressions for the regions $0<\d<0.5$, $1.1<\d<1.4$ and
$1.7<\d<2.0$. 
 
\end{section}
%
%
\begin{section}*{V. CONCLUSIONS}

We now discuss the nature of the transition as we go from zero to high
temperatures. In quantum mechanics, definitive criteria for the
continuity or discontinuity (corresponding to second order and first
order respectively) in the
derivative of the action have been obtained by Chudnovsky
\cite{chudnovsky} and Garriga \cite{garriga}. It has further been
shown that the lowest action at any temperature is possessed by either
the zero temperature or the high temperature solutions.

 In quantum field theory the situation seems to be
different. Both Ferrera \cite{ferrera} and we \cite{hatem} find that
there is an interpolating solution which can be used to determine
whether the transition is first order or second order (i.e. with or
without a kink).

 In Sec. III we find that for a thin wall ($\d \approx 2$) the
interpolating solution has a singularity at $\beta=\pi R$. The
expansion in terms of $y_\l$ breaks down at this point. So we do not
expect a real singularity at this point. However our numerical
solutions (as well as those of Ferrera) show that a kink is present in
the TWA, showing that the transition is first order. For $\d \approx
0$ (thick wall) we find that there is no kink and the transition is
smooth (second order).

 It seems, therefore, a reasonable conclusion that below $\d <1.2$
($y_\l \approx 0$) we have a second order transitions and the graph of the
action against $\beta$ is
smooth. For $\d >1.2$ we have a kink and the transition is first
order. Thus for a potential with a $\vp$ symmetry breaking and
coupling $f=0.75$ corresponding to $\d=0.65$, Ferrera finds a smooth
transition. With a $\vp^3$ symmetry breaking term and $f=0.75$
corresponding to $\d=1.46$ (see table III in our paper \cite{hatem})
we find a kink. So by transforming the potential to the Adams form and
looking at 
the resulting $\d$ we can predict whether there will be a first or a
second order transition. 

 It is suggested that our method could be used to study in detail the
nature of the phase transition in electroweak theory. Such a study
could be of importance in models of electroweak baryogenesis and other
phenomena in the early universe.

\end{section}

%
\newpage
\begin{appendix}

\section{EXPRESSIONS OF ELLIPTIC INTEGRAL IN TERMS OF THE BASIC
INTEGRALS $E_0$ AND $E_1$} \label{app1}
For $\ka <1$
\bea
E_0 & = & \int_0^1 \frac{dt}{\sqrt{1-t^2}\sqrt{1-\ka^2 t^2}}
\\ [0.3cm]
E_1 & = & \int_0^1 \frac{dt\sqrt{1-\ka^2 t^2}}{\sqrt{1-t^2}}
\\ [0.3cm]
E_3 & = & \int_0^1 \frac{dt(1-\ka^2 t^2)^{3/2}}{\sqrt{1-t^2}}
=E_1(\frac{4}{3}-\frac{2}{3}\ka^2) + E_0(\frac{\ka^2-1}{3})
\\ [0.3cm]
E_T & = & \int_0^1 \frac{dt\sqrt{1-\ka^2 t^2} t^2
(1-t^2)}{\sqrt{1-t^2}} =
\frac{2E_1}{15}(\frac{1}{\ka^4}-\frac{1}{\ka^2}+1) +
\frac{E_0}{15}(-\frac{2}{\ka^4}+\frac{3}{\ka^2}-1) 
\\ [0.3cm]
E_1'& = &\frac{E_1-E_0}{2\ka^2}
 \\[0.3cm]
E_3' & = & \frac{E_0}{2}(1-\frac{1}{\ka^2}) + E_1(-1+\frac{1}{2\ka^2})
 \\ [0.3cm]
E_T' & = &
\frac{E_1}{15}(-\frac{4}{\ka^6}+\frac{3}{2\ka^4}+\frac{1}{\ka^2})
+ \frac{E_0}{15}(\frac{4}{\ka^6}-\frac{7}{2\ka^4}-\frac{1}{2\ka^2}) 
\eea
For $\ka > 1$
\bea
E_0(1/\ka^2) & =&  \int_0^1
\frac{dt}{\sqrt{1-t^2}\sqrt{1-t^2/\ka^2}} 
 \\ [0.3cm]
E_1(1/\ka^2) &  = & \int_0^1 \frac{dt\sqrt{1-t^2/\ka^2}}
{\sqrt{1-t^2}} 
 \\ [0.3cm]
E_{04}(\ka^2) & =&  \int_0^{1/\ka} \frac{dt}{\sqrt{1-t^2}\sqrt{1-\ka^2
t^2}} 
\\ [0.3cm]
E_{14}(\ka^2)&  = & \int_0^{1/\ka} \frac{dt\sqrt{1-\ka^2
t^2}}{\sqrt{1-t^2}} = \ka E_1(1/\ka^2)-\frac{\ka^2-1}{\ka}
E_0(1/\ka^2) 
\\ [0.3cm]
E_{34}(\ka^2) & = & \int_0^{1/\ka}
\frac{dt(1-\ka^2t^2)^{3/2}}{\sqrt{1-t^2}}
=\frac{1}{\ka}\Bigg[E_1(1/\ka^2)\Big(\frac{4\ka^2}{3}-
\frac{2\ka^4}{3}\Big) 
\nonumber \\ 
& + & E_0(1/\ka^2)\Big(1-\frac{5\ka^2}{3}
 +  \frac{2\ka^4}{3}\Big) \Bigg]
\\ [0.3cm]
E_{T4} (\ka^2)& =&  \int_0^{1/\ka}  \frac{dt\sqrt{1-\ka^2 t^2} t^2
(1-t^2)}{\sqrt{1-t^2}} = \frac{1}{\ka} \Bigg[ E_1(1/\ka^2)
\Big(-\frac{2}{15}+\frac{2}{15\ka^2}+\frac{2\ka^2}{15} \Big ) 
\nonumber \\ 
& + & E_0(1/\ka^2) \Big
(\frac{1}{5}-\frac{1}{15\ka^2}-\frac{2\ka^2}{15} 
\Big) \Bigg]
\\ [0.3cm]
\frac{dE_0(1/\ka^2)}{d\ka^2} & = & \frac{1}{2\ka^2}E_0(1/\ka^2) -
\frac{E_1(1/\ka^2)}{2(\ka^2-1)} 
\\ [0.3cm]
\frac{dE_1(1/\ka^2)}{d\ka^2} & = &
\frac{1}{2\ka^2}\Bigg(E_0(1/\ka^2)-E_1(1/\ka^2)\Bigg) 
\\ [0.3cm]
\frac{dE_{14}(\ka^2)}{d\ka^2} & = &
\frac{1}{2\ka}\Bigg(E_1(1/\ka^2)-E_0(1/\ka^2)\Bigg) 
\\ [0.3cm]
\frac{dE_{34}(\ka^2)}{d\ka^2} & = &
E_(1/\ka^2)\Bigg(\frac{1}{2\ka}-\ka\Bigg)+
E_0(1/\ka^2)\Bigg(-\frac{1}{\ka}+k\Bigg) 
\\ [0.3cm]
\frac{dE_{T4}(\ka^2)}{d\ka^2} & = & E_0(1/\ka^2)
\Bigg(\frac{2}{15\ka^5}-\frac{1}{15\ka^3}-\frac{1}{15\ka} \Bigg) +
E_1(1/\ka^2)\Bigg(-\frac{4}{15\ka^5}+\frac{1}{10\ka^3}+\frac{1}{15\ka}
\Bigg) 
\eea
%
%
\section{EVALUATION OF INTEGRALS INVOLVING THE FERMI FUNCTION}
\label{app2}
For large $R^2/\l^2$ an asymptotic expansion may be obtained
through a method due to Sommerfeld as follows. Let
$y=\rho^2/\l^2$ and $y_0=R^2/\l^2$, then 
\bea
\in \frac{dy y^{1/2}}{(\z)^4} & =& \frac{2}{3} y_0^{3/2} -
\frac{11}{6} 
y_0^{1/2} + (\frac{\pi^2}{12}+\frac{1}{2}) y_0^{-1/2} + O(y_0^{-3/2})
\\ [0.3cm]
\in \frac{dy y^{1/2}}{(\z)^3} & = & \frac{2}{3} y_0^{3/2} -
\frac{3}{2} 
y_0^{1/2} + (\frac{\pi^2}{12}+\frac{1}{4}) y_0^{-1/2} + O(y_0^{-3/2})
\\ [0.3cm]
\in \frac{dy y^{1/2}}{(\z)^2} & = & \frac{2}{3} y_0^{3/2} - 
y_0^{1/2} + \frac{\pi^2}{12} y_0^{-1/2} + O(y_0^{-3/2})
\\ [0.3cm]
\in \frac{dy y^{1/2}}{\z} & = & \frac{2}{3} y_0^{3/2} +
\frac{\pi^2}{12} y_0^{-1/2} + O(y_0^{-3/2})
\\ [0.3cm]
\in \frac{dy y^{1/2}\zz}{(\z)^4} & = & \frac{1}{6} y_0^{1/2}
+ O(y_0^{-3/2})
\\ [0.3cm]
\in \frac{dy y^{3/2}\zz}{(\z)^4} & = & \frac{1}{6} y_0^{3/2} +
(\frac{\pi^2}{48}-\frac{1}{8}) y_0^{-1/2} + O(y_0^{-3/2})
\eea

%
\end{appendix}
%
\newpage
\bibliography{plain}

%
\newpage
\begin{section}*{Figure Caption}
\begin{enumerate}

\item[] FIG. 1. Temperature dependence of the Euclidean action in
thin-wall limit: $S(T)$ vs $\beta$ for $\d=1.90$.

\item[] FIG. 2. Temperature dependence of the Euclidean action in
thin-wall limit: $S(T)$ vs $\beta$ for $\d=1.85$.

\item[] FIG. 3. Temperature dependence of the Euclidean action in
thick-wall limit: $S(T)$ vs $\beta$ for $\d=0.30$.

\item[] FIG. 4. Temperature dependence of the Euclidean action in
thick-wall limit: $S(T)$ vs $\beta$ for $\d=0.10$.
\end{enumerate}
\end{section}

\newpage
\begin{figure}[ht]
\vskip 15truecm

\includegraphics{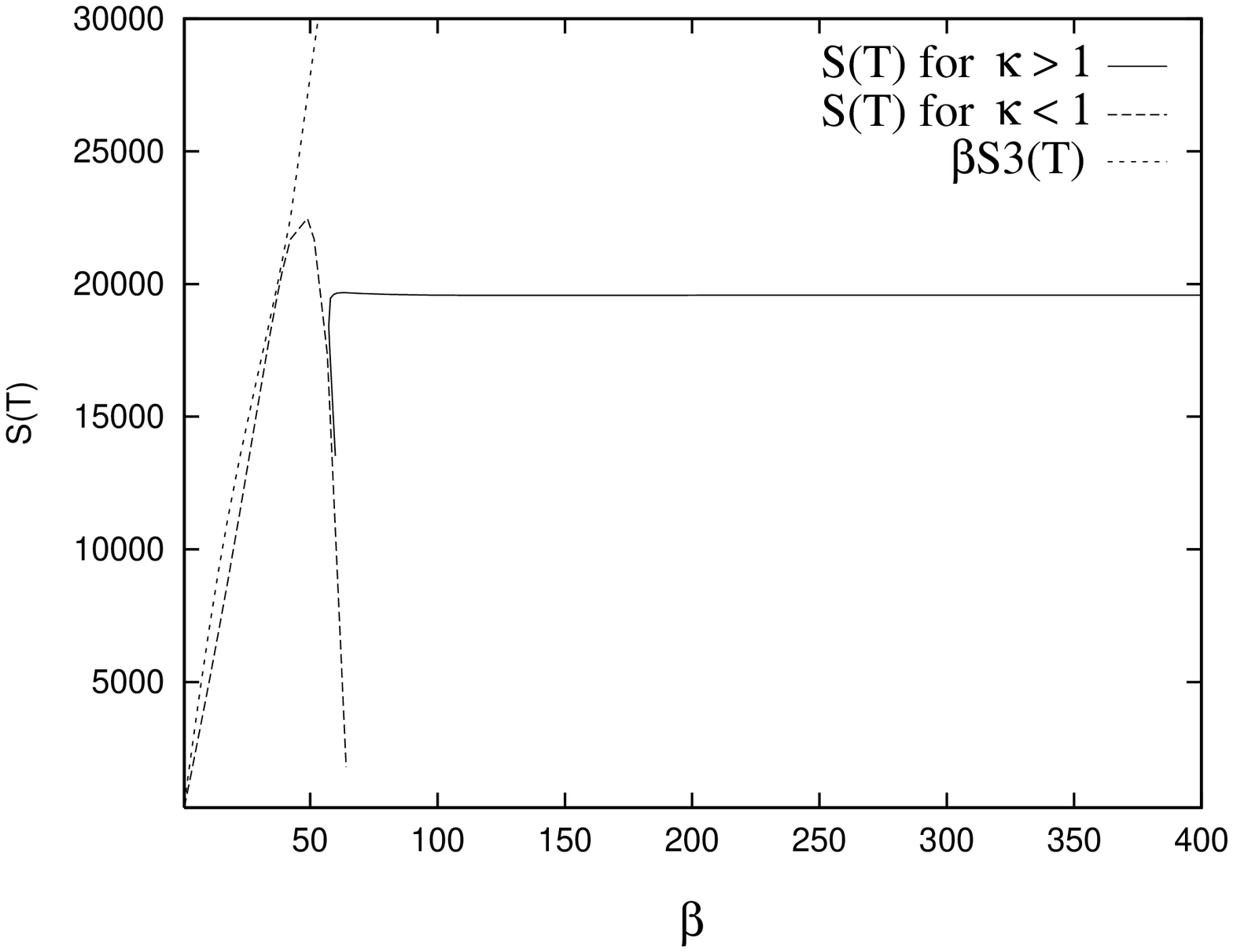} 

\caption{}
\end{figure}
\newpage
\begin{figure}[ht]
\vskip 15truecm

 \includegraphics{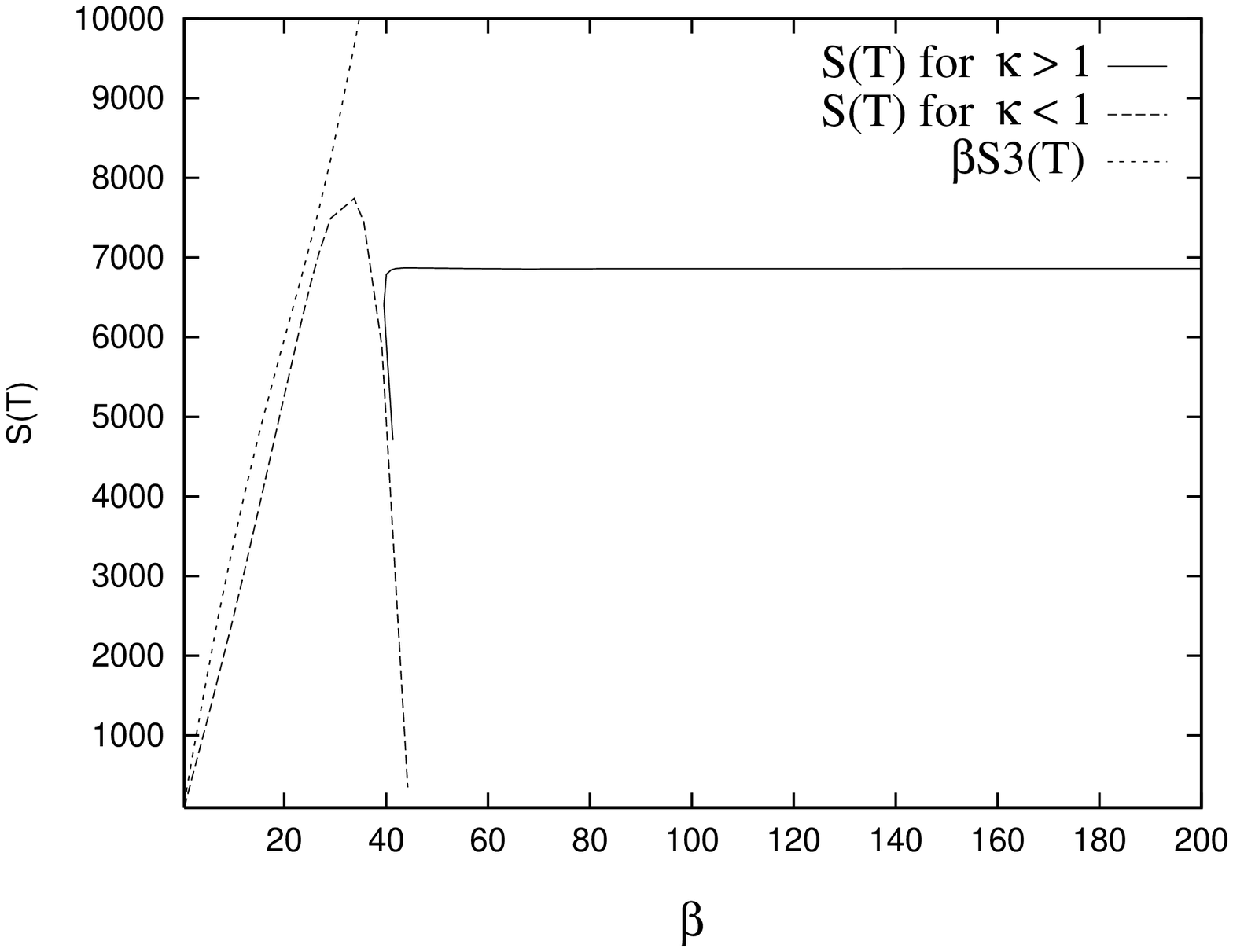} 

\caption{}
\end{figure}
\newpage
\begin{figure}[ht]
\vskip 15truecm

 \includegraphics{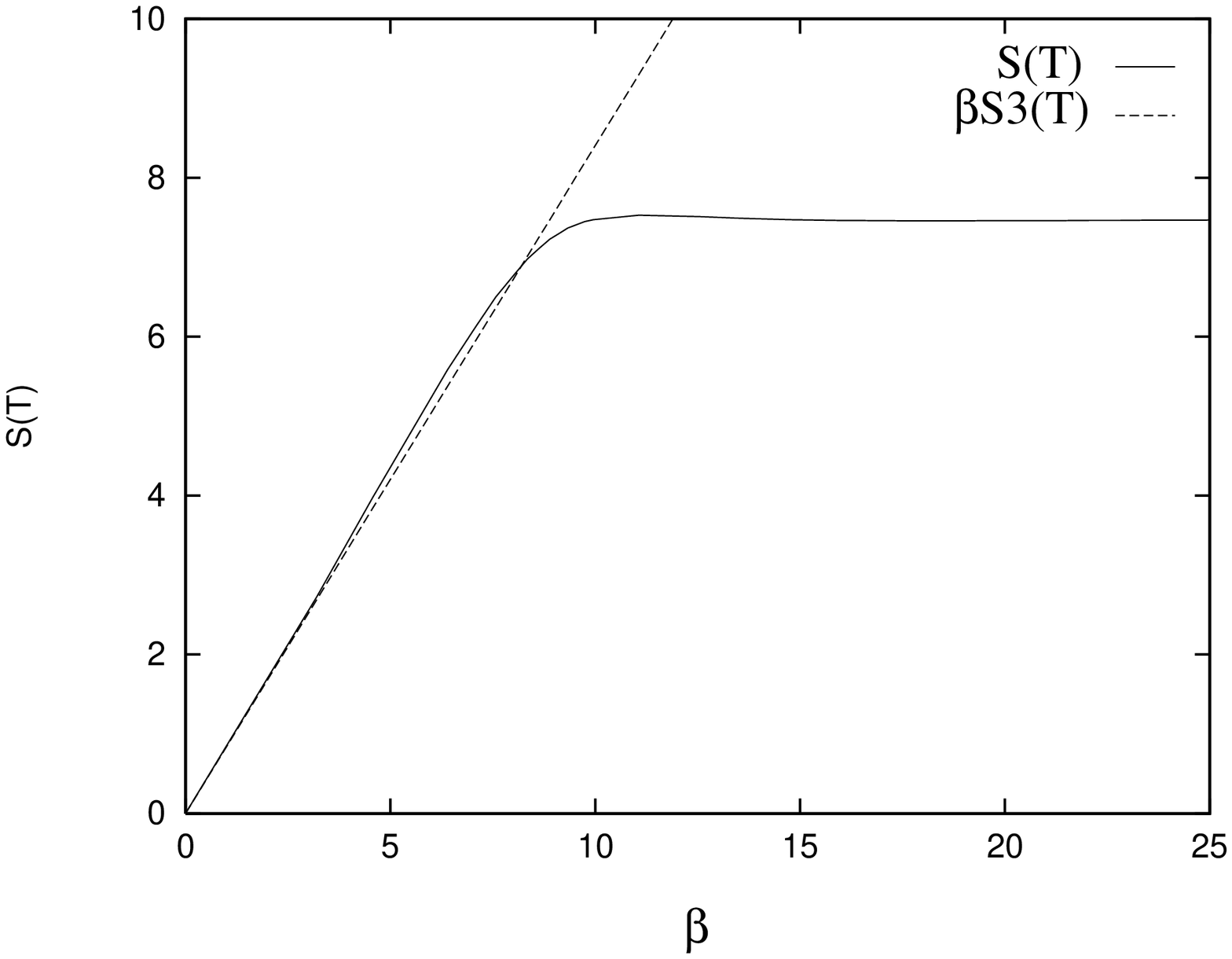} 

\caption{}
\end{figure}
\newpage
\begin{figure}[ht]
\vskip 15truecm

 \includegraphics{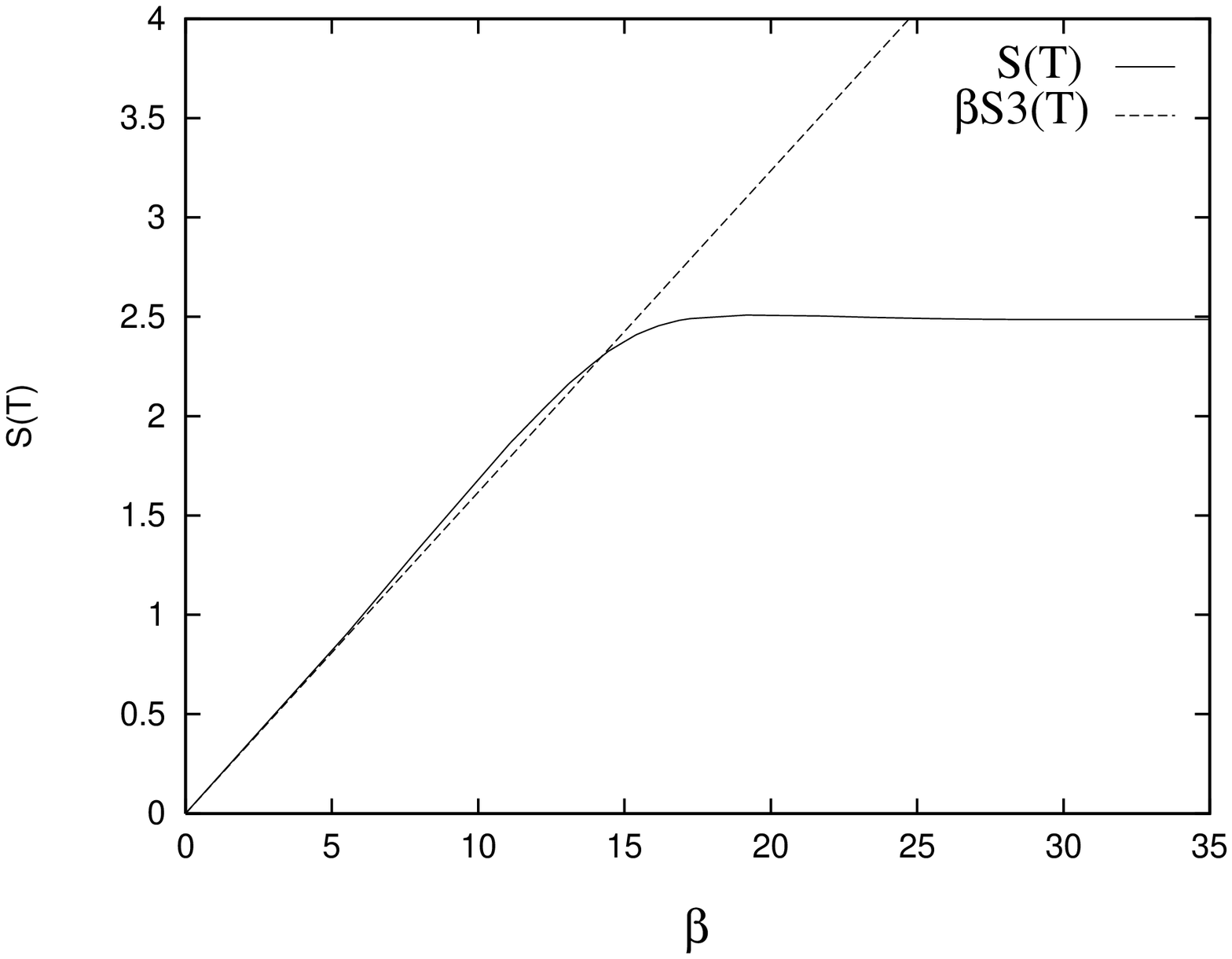} 

\caption{}
\end{figure}
\end{document}